\documentclass[twocolumn,preprintnumbers,amsmath,amssymb]{revtex4}
\begin{document}
\preprint{UH-511-1047-04}
\title{Supersymmetry breaking as the origin of flavor}
\author{Javier Ferrandis}
\email{ferrandis@mac.com}
\homepage{http: // homepage.mac.com / ferrandis}
\affiliation{Department of Physics \& Astronomy\\
 University of Hawaii at Manoa\\
 2505 Correa Road\\
 Honolulu, HI, 96822  }
\author{Naoyuki Haba}
\email{haba@ias.tokushima-u.ac.jp}
\affiliation{Institute of theoretical Physics \\
 University of Tokushima\\
 Tokushima 770-8502, Japan } 
\begin{abstract}
We present an effective flavor 
model for the radiative generation of
fermion masses and mixings based on a ${\rm SU(5)}_{\rm V}\times {\rm U(2)}_{\rm H}$ symmetry.
We assume that the original source of flavor breaking resides in
the supersymmetry breaking sector. Flavor violation
is transmitted radiatively to the fermion Yukawa couplings
at low energy through finite supersymmetric threshold corrections.
This model can fit the fermion mass ratios and CKM matrix
elements, explain the non-observation of proton decay, and
overcome the present constraints on flavor changing processes
through an approximate radiative alignment between the Yukawa and the 
soft trilinear sector.
The model predicts relations between dimensionless fermion
mass ratios in the three fermion sectors, and the quark mixing angles,
$ \left|V_{us}\right|  \approx 
\left[ \frac{m_{d}}{m_{s}} \right] ^{\frac{1}{2}}
\approx \left[ \frac{m_{u}}{m_{c}} \right]^{\frac{1}{4}} 
\approx 3 \left[ \frac{m_{e}}{m_{\mu}} \right]^{\frac{1}{2}}$ and 
$\frac{1}{2} \left|\frac{V_{cb}}{V_{us}}\right|  \approx \left[ \frac{m_{s}^{3}}{m_{b}^{2}m_{d}} \right] ^{\frac{1}{2}}
\approx \left[ \frac{m_{c}^{3}}{m_{t}^{2}m_{u}} \right] ^{\frac{1}{2}}
\approx  \frac{1}{9} \left[ \frac{m_{\mu}^{3}}{m_{\tau}^{2}m_{e}} \right]^{\frac{1}{2}}$,
which are confirmed by the experimental measurements. 
\end{abstract}
\maketitle
\newpage
%
\section{Introduction}
It is commonly assumed that the flavor mixing in the 
supersymmetric limit of the Minimal Supersymmetric Standard Model ({\it hereafter} MSSM)
is the same as in the Standard Model.
Accordingly most of the supersymmetric ({\it hereafter }SUSY) models of flavor
proposed to date have tried to explain the
fermion mass hierarchies by breaking flavor symmetries 
in the superpotential. This explanation, however, comes easily into conflict
with the present experimental constraints on flavor changing processes.
To solve this problem, it has been proposed that there is
flavor conservation in the supersymmetry breaking sector (universality)
or alternatively that the flavor violation in the supersymmetry breaking sector
is aligned with the flavor violation in the Yukawa sector to a high degree (alignment). 
Although some flavor models have been proposed that predict universality or
alignment, such conditions are usually satisfied at a high energy scale
and are spoiled through renormalization group effects.
Flavor, when it originates in the superpotential, 
is transmitted to the soft supersymmetry 
breaking sector through renormalization
group running from the unification scale down to the electroweak scale,
easily overcoming the present constraints on flavor changing processes.
There is also a third possibility, the so-called decoupling scenario, that 
assumes that the masses of the first and second generation sfermions are heavy 
enough to suppress all flavor violating processes below present limits.

The presence of flavor violation in the superpotential,
in the particular case of supersymmetric grand unified models,
causes another problem: the existence of dimension five operators that 
accelerate proton decay, thereby ruling out minimal SUSY SU(5) models.

There is an alternative possibility, which has not received much attention 
until recently \cite{Ferrandis:2004ng}. The lighter fermion masses 
may be a higher order radiative effect
as suggested by the observed fermion mass hierarchies. This is not a new idea,
following the suggestion by S.~Weinberg \cite{Weinberg:1971nd,Weinberg:1972ws} 
of a mechanism to generate radiatively the electron mass from a tree level muon mass
several proposals were published.
The program, however, was considered more difficult to implement
in the context of supersymmetric models since,  
as pointed out by 
L.~Iba\~nez, if supersymmetry is spontaneously broken
only tiny fermion masses could be generated radiatively \cite{Ibanez:1982xg}.
On the other hand, the presence of soft supersymmetry
breaking terms allows for the radiative generation of quark and charged 
lepton masses through 
sfermion--gaugino loops.
The gaugino mass provides the violation of fermionic chirality required by a fermion
mass, while the soft breaking terms provide the violation of chiral flavor symmetry. 
This idea was suggested in 1982 by 
W.~Buchmuller and D.~Wyler \cite{wyler} and was later rediscovered 
in Refs.~\cite{Hall:1985dx,Banks:1987iu,Kagan:1987wf,Ma:1988fp}.
Additional implications of this possibility were
subsequently studied in Refs.~\cite{softRadSusy,Borzumati:1999sp,Arkani-Hamed:1995fq,su5relations},
but no complete flavor model implementing radiative generation of fermion masses has been
proposed to date.

In this paper we present a supersymmetric model, 
based on a U(2) horizontal flavor symmetry, 
that generates fermion masses radiatively.
In the context of the MSSM, 
the possibility that the quark mixing and the fermion masses 
for the first two generations can actually be generated radiatively
has been recently pointed out by one of the authors \cite{Ferrandis:2004ng},
in this paper we continue that investigation.
In the U(2)-flavor model here presented 
flavor breaking originates in the supersymmetry breaking sector
and is transmitted radiatively to the fermion sector 
at low energy as mentioned above.
It is the main point of this paper to show that 
supersymmetry breaking models of flavor exist that not only can fit the
fermion mass ratios and the quark mixing angles, but also offer an alternative 
solution to the SUSY flavor and proton decay problems.
The basic conditions that we expect from a unified
supersymmetric theory that generates fermion masses radiatively are,
\begin{enumerate}
\item
A symmetry or symmetries of the superpotential guarantee
flavor conservation and precludes tree--level 
masses for the first and second generations 
of fermions in the supersymmetric limit.
\item
The supersymmetry-breaking terms receive small corrections, 
which violate the symmetry of the superpotential and are responsible
for the observed flavor physics.
\end{enumerate}
In this case the Yukawa matrices
provided as a boundary condition for the MSSM 
at some high energy scale are of the form, 
\begin{equation}
{\bf Y}_{D,U,L} = \left[
 \begin{array}{ccc}
 0 & 0 & 0 \\
 0 & 0 & 0\\
 0 & 0 & y_{b,t,\tau}
\end{array}
\right], 
\label{Yukansatz}
 \end{equation}
where ${\bf Y}_{D,U,L}$ are the $3\times 3$ 
quark and lepton Yukawa matrices.
We will assume that supersymmetry and flavor breaking 
are linked in such a way that after supersymmetry breaking,
non-trivial flavor mixing textures are generated
in the supersymmetry breaking sector. 
It is known that in an effective field theory format 
holomorphic trilinear soft supersymmetry breaking 
terms can originate, below the supersymmetry-breaking messenger scale $M$,
in operators that couple to the supersymmetry 
breaking chiral superfields, ${\cal Z}$. 
These operators are generically of the form,
\begin{equation}
\int d^{2} \theta
\left( \frac{{\cal Z}}{M} \right)  H_{\alpha} \phi_{L} \phi_{R} +c.c
\end{equation}
where $\left<{\cal Z}\right> = \left<{\cal Z}_{s}\right> + 
\left<{\cal Z}_{a} \right> \theta^{2}$.
In general the vacuum expectation value ({\it hereafter} vev) of the
auxiliary component $\left<{\cal Z}_{a}\right>$
parametrizes the scale of supersymmetry breaking, $M_{S}^{2}$.
Flavor violation may arise only in the soft terms, for instance, 
if supersymmetry breaking superfields, ${\cal Z}$, transform
non-trivially under flavor symmetries while  
the vev of the scalar component of ${\cal Z}$ vanishes, 
$\left<{\cal Z}_{s} \right>=0$,
or is much smaller than the messenger scale. 
In this case we expect the $3\times 3$ trilinear soft supersymmetry-breaking
matrices to look like,
\begin{equation}
{\bf A}_{D,U,L} = A {\cal O}(\lambda),
\label{softansatz}
 \end{equation}
where ${\cal O}(\lambda)$ 
represents generically some 
dimensionless flavor-violating polynomial matrix 
that can be expanded in powers $\lambda$,
a flavor-breaking perturbation parameter, with $\lambda<1$. Flavor violation
is transmitted to the fermion sector, i.e. to the Yukawa couplings, 
through sfermion-gaugino loops. We require 
the magnitude of $\lambda$, which must be determined a posteriori 
by the ratios between measured fermion masses,
to be consistent with present
constraints on supersymmetric contributions to flavor changing 
processes.
Finally, we note that only finite corrections
can generate off diagonal entries in the Yukawa matrices, since
the structure of the Yukawa matrices
given by Eq.~\ref{Yukansatz} is renormalization scale independent,
in other words renormalization group running from the unification scale down
to the SUSY spectra decoupling scale cannot generate off-diagonal Yukawa couplings.
The opposite, however, is not true, {\it i.e.} flavor violation in the Yukawa matrices
would transmit to the soft supersymmetry breaking
sector through renormalization group running.
This constitutes a common problem of
all the theories that locate the origin of flavor breaking in the superpotential.
\section{A ${\rm U(1)}_{H}$ toy model}
We will start with a two-generation toy model that contains the necessary
ingredients to radiatively generate fermion masses and mixings.
The model is based on a ${\rm U(1)}_{H}$ horizontal flavor symmetry.
The particle content of the model is summarized in table~\ref{toyU1}.
It is possible to choose the charge asignments in such a way that
only one generation is allowed to have a tree level 
Yukawa coupling in the superpotential, see table~\ref{toyU1}.
In this case, the tree level $2\times 2$ Yukawa matrix is,
\begin{equation}
{\bf Y} = \left[
 \begin{array}{cc}
 0 & 0 \\
 0 & y
\end{array}
\right], 
\label{toyYukU1}
 \end{equation}
where all except the (2,2) entry are disallowed by the ${\rm U(1)}_{H}$ symmetry.
We will assume that there is one supersymmetry breaking chiral superfield, ${\cal Z}$,
carrying ${\rm U(1)}_{H}$ flavor charge. 
The field ${\cal Z}$ is a spurion whose sole role is to communicate flavor as well as
supersymmetry breaking to the matter fields. We will assume
a zero vev for the scalar component of ${\cal Z}$
but non-zero for its auxiliary component. 
This condition can be achieved through an O'Raifeartaigh
type model superpotential for the ${\cal Z}$ superfield \cite{O'Raifeartaigh:pr}.
Trilinear soft supersymmetry-breaking 
terms are generated by operators generically of the form,
\begin{equation}
\int d^{2} \theta   
\left( \frac{{\cal Z}}{M_{\rm F}} \right)  f {\cal H} {\cal L}_{a} {\cal R}_{b} +c.c
\label{toysofts}
\end{equation}
where $M_{\rm F}$ is the flavor breaking scale, 
${\cal H}$ is the Higgs superfield, and $f$ is a dimensionless flavor blind 
coupling determined by the underlying theory ($f$ is flavor blind because
our basic assumption is that the underlying exactly supersymmetric model
that generates these non--renormalizable operators is flavor conserving).
We also need a flavor singlet supersymmetry breaking
superfield, ${\cal G}$, to generate a mass for the gaugino,
$m_{\widetilde{g}}$, from 
operators of the form,
\begin{equation}
\int d^{2} \theta 
\left( \frac{{\cal G}}{M} \right) \widetilde{g} \widetilde{g} +c.c,
\label{toygauginos}
\end{equation}
where $M$ is the messenger scale. We obtain
$m_{\widetilde{g}} = \left<{\cal G}\right>/M$.
We notice that, if no symmetry inhibits it, 
${\cal G}$ could 
generate an additional contribution to the trilinear soft supersymmetry breaking terms
introduced by the operator ${\cal G}  {\cal H} {\cal L}_{2} {\cal R}_{2}$.
Assuming the charge asignments in table~\ref{toyU1},
\begin{table}
\centering    
\begin{tabular}{|c|c|c|c|c|c|c|c|c|}
\hline \hline
fields & ${\cal G}$ & ${\cal Z}$ & ${\cal L}_{1}$ & ${\cal R}_{1}$ & ${\cal L}_{2}$ & ${\cal R}_{2}$ 
& ${\cal H}$ & $\widetilde{g}$\\
\hline 
${\rm U(1)}_{H}$ & 0  & -1 & 1 & 1 & 0 & 0 & 0 & 0 \\
\hline
\end{tabular} 
\caption{\rm Particle content of a two-generation ${\rm U(1)}_{H}$ flavor toy model 
with radiative fermion mass generation. ${\cal G}$ and ${\cal F}$
are supersymmetry breaking superfields, ${\cal L}_{a}$ and ${\cal R}_{a}$  ($a=1,2$) 
are matter superfields, ${\cal H}$ is the Higgs superfield
and $\widetilde{g}$ is the gaugino superfield.}
\label{toyU1}    
\end{table} 
we obtain from Eq.~\ref{toysofts} for the soft trilinear matrix,
\begin{equation}
{\bf A} = A \left[
 \begin{array}{cc}
 0 & \lambda \\
 \lambda & 1
\end{array}
\right], 
\label{toyA}
 \end{equation}
where $A= f \left<{\cal G}\right> /M $ and  $\lambda =  
\left<{\cal Z}\right>/ (M_{\rm F}m_{\widetilde{g}}) $.
We will assume that $\lambda \lesssim {\cal O}(1)$.
The parameter $\lambda$ determines the magnitude of flavor violation. 
Additionally, soft mass matrices are generated
from operators generically of the form,
\begin{equation}
\int d^{4} \theta 
\left( \frac{{\cal Z} {\cal Z}^{\dagger}}{M_{\rm F}^{2}} + 
\frac{{\cal G} {\cal G}^{\dagger}}{M^{2}} +
\frac{\rho}{M M_{\rm F}} \left( {\cal G} {\cal Z}^{\dagger} +
{\cal Z} {\cal G}^{\dagger} \right)
\right) 
k^{2 } \phi^{\dagger} \phi,
\label{toysoftsmass}
\end{equation}
where $k^{2}$ is a dimensionless coefficient determined by the underlying theory.
For the case study in table~\ref{toyU1}, we obtain the $2 \times 2$ 
soft mass matrices,
\begin{equation}
\widetilde{{\bf M}}_{\cal L}^{2}= \widetilde{{\bf M}}_{\cal R}^{2} = \widetilde{m}^{2} \left[
 \begin{array}{ccc}
 1 & \rho \lambda \\
\rho \lambda & 1  
\end{array}
\right], 
\label{toysoftMD}
 \end{equation}
where $\widetilde{m} = k \sqrt{1+\lambda^{2}} \left<{\cal G}\right>/M$.
We note that if we did not allow for mixing between flavor violating
and flavor conserving supersymmetry breaking fields, {\it i.e.} $\rho=0$,
there would be no flavor mixing in the soft mass matrices.
In the presence of flavor violation in the soft sector, the left and right handed components
of the sfermions mix. For instance, in the gauge basis 
the $4\times 4$ sfermion mass matrix is given by, 
\begin{equation}
{\cal M}^{2} = 
\left[
\begin{array}{cc}
\widetilde{\bf M}^{2}_{\cal L} + v^{2} {\bf Y}^{\dagger}{\bf Y}
 & {\bf A}^{\dagger} v \\
 {\bf A} v   & \widetilde{\bf  M}^{2}_{\cal R} +
 v^{2} {\bf Y}{\bf Y}^{\dagger}
 \\
\end{array}
\right],
\end{equation}
where $v= \left<{\cal H }\right>$, 
$\widetilde{\bf M}^{2}_{\cal L}$ and $\widetilde{\bf M}^{2}_{\cal R}$
are the $2 \times 2$ right handed and left handed soft mass matrices given above
(including D-terms), ${\bf A}$ is the $2 \times 2$ soft trilinear matrix and
${\bf Y}$ is the $2 \times 2$ tree level Yukawa matrix.
${\cal M}^{2}$ is diagonalized by a $4 \times 4$ unitary matrix, 
${\cal D}$.  In general, the dominant finite 
one--loop contribution to the $2 \times 2$ Yukawa matrix
is given by the gaugino-sfermion loop,
\begin{equation}
({\bf Y})_{ab}^{\rm{rad}} = \frac{\alpha }{\pi} 
m_{\widetilde{g}} 
\sum_{c} {\cal D}_{ac} {\cal D}_{(b+2) c}^{*} 
B_{0}(m_{\widetilde{g}}, m_{\widetilde{f}_{c}}) 
\end{equation}
where $\widetilde{f}_{c}$ ($c =1, \cdot \cdot \cdot, 4$)
are sfermion mass eigenstates, $m_{\widetilde{g}}$ is the gaugino mass,
and $\alpha$ is the gauge coupling of the theory.
$B_{0}$ is a known function defined in the appendix. We observe that
the contributions from left and 
right handed sfermion mixings involving the soft masses are much smaller.
For simplicity we will assume from now on that there are no CP-violating phases
in the soft parameters. 
Moreover, as a consequence of the approximate 
sfermion mass degeneracy predicted by the model when $\lambda \ll 1$,
the dominant contribution to the radiatively generated Yukawa couplings
can be simply expressed as,
\begin{equation}
{\bf Y}^{\rm{rad}} = \frac{2 \alpha }{ \pi} 
m_{\widetilde{g}} {\bf A} 
F(\widetilde{m}_{f}, \widetilde{m}_{f},m_{\widetilde{g}}),
\end{equation}
where the function $F$ is given in the appendix.
We then obtain a simple expression for 
the radiatively corrected mass matrix,
\begin{equation}
{\bf m} =
v \left(  {\bf Y} +{\bf Y}^{\rm{rad}} \right)=
 m \left[
 \begin{array}{cc}
 0 & \gamma \lambda \\
 \gamma \lambda & 1 
\end{array}
\right], 
\label{toymD}
 \end{equation}
where $m$ and $\gamma$ are given by,
\begin{equation}
m = y v \left( 1  + \zeta  \right), \quad \gamma = \frac{\zeta}{1+\zeta},
\end{equation}
and
\begin{equation}
\zeta  = 
\frac{2 \alpha }{\pi y } m_{\widetilde{g}} A F(\widetilde{m}, \widetilde{m},m_{\widetilde{g}}).
\end{equation}
The loop factor $\gamma$ encodes the dependence on the SUSY spectra and
parametrizes the breaking of the alignment 
between the soft trilinear and the Yukawa sectors caused 
by the presence of the tree level mass $m$.
Although not diagonal in the gauge basis, the matrix ${\bf m}$ 
can be brought to diagonal form in the mass basis by a unitary diagonalization,
$ {\cal V}^{\dagger}  {\bf m} {\cal V} =  \left( m_1, m_{2} \right)$.
Therefore the one-loop mass matrix of our toy model 
makes the following prediction for the mass ratio between the
radiatively generated mass, $m_{1}$, and the tree level one,
\begin{equation}
\frac{m_{1}}{m_{2}} =  \gamma^{2} \lambda^{2}.
\label{toyrats}
\end{equation}
The flavor mixing is given by a
CKM-like mixing matrix,
\begin{equation}
{\cal V} = \left[
 \begin{array}{cc}
 1 - \frac{1}{2} \gamma^{2} \lambda^{2} & \gamma \lambda \\
  -\gamma \lambda & 1 - \frac{1}{2} \gamma ^{2}\lambda^{2}
\end{array}
\right], 
\label{toyVCKM}
 \end{equation}
We note that the mass ratio and the flavor mixing are determined
by two basic parameters of the model, the flavor breaking parameter $\lambda$
and the loop suppression factor $\gamma$. Moreover,
the mass ratio is directly correlated with the flavor mixing angle, ${\cal V}_{12}$, 
\begin{equation}
\frac{m_{1}}{m_{2}} =  \gamma^{2} \lambda^{2} = {\cal V}_{12}^{2}
\label{toyrats}
\end{equation}
If we could determine experimentally the mixing angle 
we could predict $m_{1}$ or viceversa.
This toy model illustrates
the mechanism that will be used in a realistic model
in the next section.
\section{${\rm U(2)}_{H}$ flavor symmetry}
In this section we will consider a realistic three generation model
based in a horizontal ${\rm U(2)}_{H}$ symmetry \cite{bhrr}. 
We will assume the usual MSSM particle content where 
third generation matter superfields,
\begin{equation}
 {\cal Q}_{3}, {\cal D}_{3}, {\cal U}_{3}, {\cal L}_{3}, {\cal E}_{3},
\end{equation}
and up and down electroweak Higgs superfields,
${\cal H}_{u}$ and ${\cal H}_{d}$,
are singlets under ${\rm U(2)}_{H}$.
We will denote them abbreviately by $\phi^{L,R}$. Let us assume that
first and second generation left and right handed superfields transform 
as contravariant vectors under ${\rm U(2)}_{H}$,
\begin{eqnarray}
\Psi_{\cal Q} &=& \left(
 \begin{array}{c}
 {\cal Q}_{1}\\
{\cal Q}_{2}
\end{array}
\right),
\Psi_{\cal U} = \left(
 \begin{array}{c}
 {\cal U}_{1}\\
 {\cal U}_{2}
\end{array}
\right),
\Psi_{\cal D} = \left(
 \begin{array}{c}
 {\cal D}_{1}\\
 {\cal D}_{2}
\end{array}
\right), \\
\Psi_{\cal L} &=& \left(
 \begin{array}{c}
 {\cal L}_{1}\\
 {\cal L}_{}{2}
\end{array}
\right),
\Psi_{\cal E} = \left(
 \begin{array}{c}
 {\cal E}_{1}\\
 {\cal E}_{2}
\end{array}
\right), 
\end{eqnarray}
We will denote them abbreviately by $\Psi_{a}^{L,R}$.
We will introduce a set of supersymmetry breaking 
chiral superfields,
\begin{equation}
{\cal S}^{ab}, \quad {\cal A}^{ab}, \quad {\cal F}^{a} \quad (a,b=1,2),
\end{equation}
that transform covariantly as a symmetric tensor, 
antisymmetric tensor and vector under ${\rm U(2)}_{H}$
(with a U(1) charge opposite to that of the matter doublets).
We will assume that at the mimimum only the auxiliary components of the 
flavor breaking superfields are non zero,
The most general form for the vevs of the flavor breaking fields is,
\begin{eqnarray}
<{\cal S}> &=& \left(
 \begin{array}{cc}
v_{\cal S}  &  0 \\
 0 & {\cal V}_{\cal S}
\end{array}
\right) \theta^{2}
\label{Svev}
, \\ 
<{\cal A}>&=& \left(
 \begin{array}{cc}
 0  &  {\cal V}_{\cal A} \\
 -{\cal V}_{\cal A} &  0
\end{array}
\right) \theta^{2}, 
\label{Avev} \\ 
<{\cal F}>  &=& \left(
 \begin{array}{c}
 v_{\cal F} \\
 {\cal V}_{\cal F}
\end{array}
\right) \theta^{2},
\label{Fvev}
\end{eqnarray}
We will assume the following particular hierarchy in the flavor breaking vevs:
$v_{\cal S} \ll {\cal V}_{\cal S}$ and for practical purposes
$v_{\cal S}= 0$. We also assume that ${\cal V}_{\cal S} = {\cal V}_{\cal F}$ and 
\begin{equation}
\left( v_{\cal F}, {\cal V}_{\cal A},{\cal V}_{\cal F}\right) =
 \left( \lambda^{2} , \lambda ^{2}, \lambda \right) M_{\rm F} \widetilde{m}.
 \label{vavf}
\end{equation}
These ad-hoc assumptions will prove a posteriori 
to be very succesfull in reproducing fermion masses.
Here $\lambda$ is the flavor breaking perturbation parameter,
$M_{\rm F}$ is the flavor breaking scale.
The only couplings allowed in the superpotential
by the ${\rm U(2)}_{H}$ horizontal symmetry
and the ${\rm SU(3)}_{C} \times {\rm SU(2)}_{L} \times {\rm U(1)}_{Y}$ vertical symmetry
are the third generation ones and the so called $\mu$--term, 
\begin{equation}
\lambda_{t} {\cal Q}_{3} {\cal U}_{3}{\cal H}_{u} +
\lambda_{b} {\cal Q}_{3} {\cal D}_{3}{\cal H}_{d} +
\lambda_{\tau} {\cal L}_{3} {\cal E}_{3}{\cal H}_{d} + \mu {\cal H}_{u} {\cal H}_{d}.
\label{superpo}
\end{equation}
We note that, in principle, two other couplings,
could be allowed in the superpotential:
${\cal L}_{3}{\cal H}_{u}$ and ${\cal Q}_{3}{\cal L}_{3}{\cal D}_{3}$. 
There are different ways to remove this unwanted couplings.
They could be forbidden imposing total fermion number conservation.
Alternatively one could impose $R$--parity conservation
defined as $R = (-)^{3B + L +2S}$, where B is the barionic number, L the leptonic number
and S the spin. A third possibility would be to extend the ${\rm U(2)}_{H}$ symmetry
to the maximal ${\rm U(3)}_{H}$ horizontal symmetry. The breaking of the ${\rm U(3)}_{H}$ symmetry 
in the direction of the third generations would leave us with our ${\rm U(2)}_{H}$ symmetry,
in such a case this bilinear interaction would not be a ${\rm U(3)}_{H}$ singlet. 
Therefore, at tree level Yukawa matrices are generically of the form,
\begin{equation}
{\bf Y} = \left[
 \begin{array}{ccc}
0 & 0 & 0 \\
0 & 0 & 0 \\
 0 & 0 & y
\end{array}
\right], 
\label{Yukrad}
 \end{equation}
First we need to introduce a flavor--singlet chiral superfield, ${\cal G}$,
to give masses to gauginos from operators of the form,
\begin{equation}
\int d^{2} \theta 
\left( \frac{{\cal G}}{M} \right) \widetilde{g} \widetilde{g} +c.c,
\label{gauginosmass}
\end{equation}
where $M$ is the messenger scale.
The gaugino mass generated is given by 
$m_{\widetilde{g}} = \left<{\cal G}\right>/M$.
Additionally, trilinear soft supersymmetry 
breaking terms are generated by operators generically of the form,
\begin{eqnarray}
\sum_{{\cal Z}={\cal S}, {\cal A} } \frac{1}{M_{\rm F}}
\int d^{2} \theta   {\cal Z}^{ab}\Psi^{L}_{a} \Psi^{R}_{b} {\cal H}_{\alpha}
 +c.c \\
\frac{1}{M_{\rm F}} \int d^{2} \theta 
\left(  \phi^{R} {\cal F}^{a} \Psi^{L}_{a} +  \phi^{L} 
{\cal F}^{a} \Psi^{R}_{a} \right) {\cal H}_{\alpha}  +c.c 
\label{softs}
\end{eqnarray}
where $M_{\rm F}$ is the flavor breaking scale,
$a=1,2$ and ${\cal H}_{\alpha}, ~\alpha=u,d$
represents any of the Higgs superfields, $\phi^{L,R}$ stands generically for
any of the right or left handed flavor--singlet matter superfields,
and $\Psi^{L,R}$ stands generically for
any of the right or left handed flavor--vector matter superfields.
The flavor singlet superfield responsible for generating gaugino masses, 
${\cal G}$, will couple to matter fields. Additionally, to add more
generality to our analysis,  
we will assume that there
could be another flavor-singlet, ${\cal J}$, which couples to the
matter superfields but does not couple to the gaugino superfields.
In the most general case there could be two additional operators 
generating soft trilinears,
\begin{equation}
\frac{1}{M}
\int d^{2} \theta 
\left( \kappa {\cal G} + \eta {\cal J} \right) \phi^{L} \phi^{R} {\cal H}_{\alpha} +c.c.,
\label{GJtril}
\end{equation}
where $\kappa$ and $\eta$ are dimensionless couplings determined
by the underlying theory. We define the soft breaking mass generated
for the ${\cal J}$ field as $m_{\widetilde{J}} = \left<{\cal J}\right>/M$.
Soft supersymmetry breaking mass matrices are
generated by operators generically of the form,
\begin{eqnarray}
 \frac{1}{M^{2}_{\rm F}}
\int  d^{4} \theta  \left(
 {\cal Z}^{\dagger}_{ac} {\cal Z}^{cb}  (\Psi^{\dagger})^{a} \Psi_{b} + 
 {\cal F}^{\dagger}_{a} {\cal F}^{b} (\Psi^{\dagger})^{a} \Psi_{b} \right) + \nonumber \\
\frac{1}{M^{2}}
\int  d^{4} \theta   ( \kappa^{\prime 2}{\cal G}^{\dagger} {\cal G} + \eta^{\prime 2} {\cal J}^{\dagger} {\cal J})
\left( (\Psi^{\dagger})^{a} \Psi_{a}  + \phi^{\dagger} \phi  \right) + \nonumber \\
\frac{1}{M M_{\rm F}}
\int  d^{4} \theta  \rho \left( ( \kappa^{\prime }{\cal G}^{\dagger}+ \eta^{\prime} {\cal J}^{\dagger} )
 {\cal F}^{b} \phi^{\dagger} \Psi_{b}  + {\rm h.c.} \right)~~~~~~~ 
\label{softs}
\end{eqnarray}
where ${\cal Z}={\cal S}, {\cal A}$. When including the last term
in the previous equation we assume that the underlying theory 
allows the flavor breaking fields to couple with the flavor singlets
${\cal G}$ and ${\cal J}$, if this were not possible $\rho=0$.
After ${\rm U(2)}_{H}$ flavor breaking the following soft trilinear matrices 
are generated, 
\begin{equation}
{\bf A} = A \left[
 \begin{array}{ccc}
0 &  \sigma \lambda^{2}  & \sigma \lambda^{2}   \\
-  \sigma \lambda^{2}  & \sigma \lambda & \sigma  \lambda \\
\sigma \lambda^{2}   & \sigma \lambda & 1
\end{array}
\right], 
\label{Asoft}
 \end{equation}
where,
\begin{equation}
A =  \left( \kappa ~m_{\widetilde{g}}  + 
\eta  ~ m_{\widetilde{J}} \right)  
\end{equation}
and the dimensionless parameter $\sigma$ is defined by,
\begin{equation}
\sigma = \frac{\widetilde{m}}{A}, 
\label{sigma}
\end{equation}
where $\widetilde{m}$ was defined in Eq.~\ref{vavf}.
After ${\rm U(2)}_{H}$ flavor breaking the following soft mass matrices 
are also generated, 
\begin{equation}
\widetilde{{\bf M}}_{L,R}^{2} = \widetilde{m}^{2}_{f} \left[
 \begin{array}{ccc}
1 + 2 \lambda^{4} \sigma^{\prime 2}& \lambda^{3} \sigma^{\prime 2} &\rho  \lambda^{2} \sigma^{\prime }  \\
\lambda^{3} \sigma^{\prime 2} & 1  + 2 \lambda^{2}\sigma^{\prime 2} & \rho \lambda \sigma^{\prime }  \\
\rho \lambda^{2} \sigma^{\prime }  & \rho \lambda \sigma^{\prime }  & 1
\end{array}
\right], 
\label{Msoft}
 \end{equation}
where, 
\begin{equation}
\widetilde{m}^{2}_{f} =  \left( \kappa^{\prime 2}  ~m^{2}_{\widetilde{g}}  +
\eta^{\prime 2}  ~ m^{2}_{\widetilde{J}} \right)  
\end{equation}
and the dimensionless parameter $\sigma^{\prime}$ is defined by,
\begin{equation}
\sigma^{\prime} = \frac{\widetilde{m}}{\widetilde{m}_{f}}. 
\label{sigmap}
\end{equation}
We note that in this scenario the amount of flavor violation
as well as the non-degeneracy predicted in the soft mass matrices 
is determined by the ratio $A/\widetilde{m}_{f}$. 
We note that
the presence of mixing between flavor breaking and flavor singlet
susy breaking fields generates flavor violating soft mass matrices
in the entries (13) and (23). 
We will see later that 
$\lambda^{2} \approx 0.05$, is approximately the Cabbibo angle squared. 
The sfermion non--degeneracy between 
first and second generations appears to order $\sigma^{\prime 2}\lambda^{2}$.

There are three interesting limits. In the first limit we assume that
there is no extra flavor singlet, ${\cal J}$,
{\it i.e.} $\eta=\eta^{\prime}=0$, and to simplify we assume $\kappa=\kappa^{\prime}=1$.
We then obtain,
\begin{eqnarray}
A &=& \widetilde{m}_{f} = m_{\widetilde{g}}, \\
\sigma &=& \sqrt{\sigma^{\prime}} = \widetilde{m}/m_{\widetilde{g}}.
\label{deltadef}
\end{eqnarray}
$\widetilde{m}$ was defined in Eq.~\ref{vavf}.
In this case all the supersymmetric spectra are correlated with the gaugino mass.
In the second limit the flavor singlet
superfield ${\cal G}$ does not couple to the matter fields, {\it i.e.} 
$\kappa=\kappa^{\prime}=0$, and to simplify we assume 
$\eta=\eta^{\prime}=1$. We then obtain,
\begin{eqnarray}
\widetilde{m}_{g} &\neq& m_{\widetilde{J}}, \\
A &=& \widetilde{m}_{f} = m_{\widetilde{J}}, \\
\sigma &=& \sqrt{\sigma^{\prime}} = \widetilde{m}/m_{\widetilde{J}},
\end{eqnarray}
where $m_{\widetilde{J}} = \left<{\cal J}\right> /M$ is in general 
different from the gaugino mass.
The third interesting limit we will consider
is specially relevant from a phenomenological point of view.
If we assume that
\begin{equation}
 A < \widetilde{m} \ll m_{\widetilde{f}},
\end{equation}
then $\sigma^{\prime} \ll 1$ and $\sigma >1$.
This case would suppress the contributions from soft masses to 
flavor violating processes while increasing the soft trilinear 
contributions to the radiatively generated Yukawa couplings.
\subsection{The down-type quark sector}
In general, 
one--loop gluino--squark exchange generates a dominant 
finite contribution to the $3 \times 3$ quark Yukawa mass matrices
given by,
\begin{equation}
({\bf Y})_{ab}^{\rm{rad}} = \frac{\alpha_{s} }{3 \pi} 
m_{\widetilde{g}} 
\sum_{c} {\cal D}_{ac} {\cal D}_{(b+3) c}^{*} 
B_{0}(m_{\widetilde{g}}, m_{\widetilde{d}_{c}}) 
\end{equation}
where $\widetilde{d}_{c}$ ($c =1, \cdot \cdot \cdot, 6$)
are squark mass eigenstates,
$\alpha_{s}$ is the strong coupling constant,
${\cal D}$ is a $6 \times 6$ down/up-type squark 
diagonalization matrix and 
$m_{\widetilde{g}}$ is the gluino mass.
The function $B_{0}$ is defined in the appendix.
For example, 
the radiatively generated $3 \times 3$ down--type quark mass 
matrix is generically of the form,
\begin{equation}
{\bf m}_{D} = <{\cal H}_{d}> ( {\bf Y} + {\bf Y}^{\rm{rad}} ) = \widehat{m}_{b} \left[
 \begin{array}{ccc}
0 &  \gamma_{b} \lambda^{2} & \gamma_{b} \lambda^{2} \\
- \gamma_{b} \lambda^{2} & \gamma_{b} \lambda &  \gamma_{b} \lambda \\
\gamma_{b} \lambda^{2} & \gamma_{b} \lambda & 1
\end{array}
\right], 
\label{YukDrad}
 \end{equation}
where $\widehat{m}_{b}$ is to first order
the running bottom mass,
\begin{eqnarray}
\widehat{m}_{b} &=& y_{b} v c_{\beta} \left( 1  + \zeta_{d} (1-  y_{b} \mu t_{\beta} /A)\right), \\
\zeta_{d}  &=& 
\frac{2 \alpha_{s} }{3 \pi y_{b} } m_{\widetilde{g}} A
 F(\widetilde{m}_{f}, \widetilde{m}_{f},m_{\widetilde{g}}).
\label{mbEq}
\end{eqnarray}
Here $t_{\beta} = \tan \beta = v_{u}/v_{d}$ 
is the ratio between the vevs of the two MSSM Higgsses 
($v _{u}=\left<{\cal H}_{u} \right>$,  $v_{d}= \left<{\cal H}_{d}\right>$),
$\mu$ is the so called $\mu$--term, introduced
in Eq.~\ref{superpo}, and 
$v = s_{W}m_{W} /\sqrt{2 \pi \alpha_{e}}= 174.5$~GeV
( $s_{W}$ is the weak mixing angle, $m_{W}$ is the SM
$W$--boson mass, and $\alpha_{e}$ is the electromagnetic
coupling constant). $\gamma_{b}$ is a loop factor that 
has a simple expression in the mass degenerate sfermion case. 
For example, in the down-type quark sector is given by,
\begin{equation}
\gamma_{b} = \frac{\zeta_{d}  \sigma }{\left( 1  + \zeta_{d} (1- y_{b} \mu t_{\beta}/A )\right)}. 
\end{equation}
Here $\sigma$ is the coefficient introduced in Eq.~\ref{Asoft};
if there is no extra flavor singlet ,${\cal J}$, 
and $\kappa=1$ $\sigma$ would be defined
as $\widetilde{m}/m_{\widetilde{g}}$. $\gamma_{b}$
encodes the dependence on the SUSY spectra and
parametrizes the breaking of the alignment 
between soft trilinear and Yukawa sectors caused by the tree level component
to the bottom mass.
We observe that there is a special limit of the previous
formulas, $y_{b} \rightarrow 0$, where the bottom quark mass
could also be generated radiatively. We will not consider that
case; we assume that all third generation fermions 
get a tree level mass.

The phenomenological implications of a mass matrix of the form 
given in Eq.~\ref{YukDrad} have been
studied by one of the authors in Ref.~\cite{Ferrandis:2004ng}; 
here we reproduce some of the results.
Although not diagonal in the gauge basis the matrix ${\bf m}_{D}$ 
can be brought to diagonal form in the mass basis by a biunitary diagonalization,
$ ({\cal V}^{d}_{L})^{\dagger} {\bf m}_{D} {\cal V}^{d}_{R}
=  \left( m_{d}, m_{s}, m_{b} \right)$.
The down--type quark mass matrix given by Eq.~\ref{YukDrad}
makes the following predictions for the quark mass ratios,
\begin{eqnarray}
\frac{m_{d}}{m_{s}} &=& \lambda^{2} ( 1 + \gamma_{b} \lambda - 2 \lambda^{2} ) + {\cal O}(\lambda^{6}) 
, \\ 
\frac{m_{s}}{m_{b}} &=&  \gamma_{b} \lambda ( 1 - \gamma_{b} \lambda + \lambda^{2} )
  + {\cal O}(\lambda^4).
 \label{downquarkrats}
 \end{eqnarray}
We can express $\lambda$ and $\gamma_{b}$
as a function of renormalization scheme and approximately
scale independent dimensionless quark mass ratios, to first order,
\begin{equation}
\lambda = \left( \frac{m_{d}}{m_{s}}\right)^{1/2}, \quad 
\gamma_{b} = \left(  \frac{m_{s}^{3}}{m_{b}^{2} m_{d}} \right)^{1/2},
\end{equation}
Using the invariant running quark mass ratios 
determined from experiment (see appendix),
we can determine $\lambda$ and $\gamma_{b}$, 
\begin{eqnarray}
\lambda &=& 0.209 \pm 0.019, 
\label{Eq:lambdad} \\
\gamma_{b} &=&  0.109 \pm 0.030. 
\label{Eq:gammab}
\end{eqnarray}
The down-type quark diagonalization matrix can be calculated 
as a function of $\lambda$ and $\gamma_{b}$;
at leading order in $\lambda$ we obtain,
\begin{equation}
\left|{\cal V}^{d}_{L}\right| =
\left[
 \begin{array}{ccc}
1 - \frac{1}{2} \lambda^{2}  &  \lambda   & 
 \gamma_{b} \lambda^{2}  \\
\lambda &  1 - \frac{1}{2} \lambda^{2} \left(1 + \gamma_{b}^{2}\right) & 
 \gamma_{b} \lambda
 \\
 \gamma_{b} \lambda^{4} 
 &  \gamma_{b} \lambda
 & 1 - \frac{1}{2} \gamma_{b}^{2} \lambda^{2}
\end{array}
\right].
\end{equation}
Using the experimentally determined values for $\gamma_{b}$ and
$\lambda$ in Eqs.~\ref{Eq:lambdad}, \ref{Eq:gammab},
we obtain the following central theoretical prediction for the $\left| {\cal V}^{d}_{L} \right|_{\rm{theo}}$ 
elements,
\begin{equation}
\left[
 \begin{array}{ccc}
0.976\pm 0.008 & 0.216\pm 0.035 & 0.0039 \pm 0.0006 \\ 
0.216\pm 0.035 & 0.974 \pm 0.007 & 0.019 \pm 0.007 \\
0.00015  & 0.019 \pm 0.007  & 0.9993 \pm 0.0001
\end{array}
\right]. 
\end{equation}
If we compare  $\left| {\cal V}^{d}_{L} \right|_{\rm{theo}}$ 
with the 90~\% C.L. 
experimental compilation of CKM matrix elements (see appendix), 
we observe that  $\left| {\cal V}^{d}_{L} \right|_{\rm{theo}}$ 
accounts quite well for the measured SM flavor violation. 
There is good agreement with 
the experimental data on CKM matrix elements 
except in the entry $\left|V_{cb}\right|$, where we observe a deficit
in the theoretical prediction, which turns out to be approximately one half of 
the measured value, {\it i.e.} $\gamma_{b} \lambda = \left|V_{cb}\right|/2$. 
We will see later that to solve this deficit we are forced to
generate half of the contribution to $\left|V_{cb}\right|$ 
from flavor violation in the up-type quark sector. 
There is a simpler alternative solution.
We can assume that the flavor mixing in the up-type quark sector
does not affect the CKM mixing matrix to leading order in $\lambda$
while the vev of the ${\cal F}$ field is instead given by, 
\begin{equation}
\left( v_{\cal F},{\cal V}_{\cal F}\right) =
 \left( \lambda^{2} , 2 \lambda \right) M_{\rm F} \widetilde{m}.
 \label{vavf2}
\end{equation}
In this case we obtain ${\bf A}_{12}={\bf A}_{21} = 2 \sigma \lambda$,
$({\bf m}_{D})_{12}=({\bf m}_{D})_{21}= 2 \gamma_{b} \lambda$ and
$\left|{\cal V}_{L}^{d}\right|= 2 \gamma_{b}\lambda$.
We also observe that the prediction for the entry $\left|(V^{d}_{L})_{31}\right|$ 
is very small, tough compatible with the measured
value of $\left|V_{td}\right|$  which carries a large uncertainity.
Before studying the up--type quark sector 
we will analyze in the next subsection the predictions
of the model for the lepton sector.
\subsection{The lepton sector and the need for $SU(5)$}
The mass matrix in Eq.~\ref{YukDrad} is very succesful in reproducing the 
down-type quark mass ratios, but it 
cannot explain correctly the measured mass ratios 
in the lepton and up--type quark sectors.
To account for the mass ratios in the lepton sector we will need to promote the 
Standard Model ${\rm SU(3)}_{c}\times {\rm U(2)}_{L} \times {\rm U(1)}_{Y}$ vertical symmetry to
the $SU(5)$ symmetry of Georgi and Glashow and assign the ${\rm U(2)}$ flavor
breaking fields to particular representations under $SU(5)$ as we will 
explain in detail below.

First, we are going to postulate
for the lepton soft trilinear matrix ${\bf A}_{L}$
a simple modification of the  
texture predicted by the minimal model in Eq.~\ref{Asoft}.
Let us assume that,
\begin{equation}
{\bf A}_{L} = A_{\tau} \left[
 \begin{array}{ccc}
0 & \sigma_{l} \lambda^{2} &  \sigma_{l} \lambda^{2}  \\
-  \sigma_{l} \lambda^{2}&  3 \sigma_{l} \lambda &  \sigma_{l} \lambda \\
 \sigma_{l} \lambda^{2}  & \sigma_{l} \lambda & 1
\end{array}
\right].
\label{ALsoft}
 \end{equation}
Here $\lambda$ was introduced in Eq.~\ref{vavf}, and
$A_{\tau}$ and $\sigma_{l}$ are coefficients analogous to the ones 
introduced in Eq.~\ref{Asoft}.
If there is no extra flavor singlet ${\cal J}$, $A$ is given by
$A_{\tau} = \kappa m_{\widetilde{\gamma}}$,
where $m_{\widetilde{\gamma}}$ is the photino mass.
We will show next that this texture
can perfectly fit the lepton mass ratios.
Later we will explain how one can obtain this texture
in a SUSY $SU(5)$ framework. 
The radiatively generated lepton Yukawa couplings are given
in this case  by,
\begin{equation}
({\bf Y}_{L})_{ab}^{\rm{rad}} = \frac{\alpha }{ 2 \pi} 
m_{\widetilde{\gamma}} 
\sum_{c} {\cal D}_{ac} {\cal D}_{(b+3) c}^{*} 
B_{0}(m_{\widetilde{\gamma}}, m_{\widetilde{l}_{c}}) 
\end{equation}
where ${\cal D}$ is the slepton $6 \times 6$ 
diagonalization matrix, $m_{\widetilde{l}_{c}}$
are slepton mass eigenvalues, and $\alpha$ is the 
running fine structure constant. 
We now obtain a simple expression for 
the radiatively corrected lepton mass matrix,
\begin{equation}
{\bf m}_{L} =  \widehat{m}_{\tau}
\left[
 \begin{array}{ccc}
 0 & \gamma_{\tau} \lambda^{2} & \gamma_{\tau} \lambda^{2}  \\
- \gamma_{\tau} \lambda^{2} &  
3 \gamma_{\tau} \lambda &  \gamma_{\tau} \lambda \\
\gamma_{\tau} \lambda^{2}  & \gamma_{\tau} \lambda & 1
\end{array}
\right],
\label{mLmat}
\end{equation}
where $\widehat{m}_{\tau}$ is to first order
the running bottom mass,
\begin{eqnarray}
\widehat{m}_{\tau} &=& y_{\tau} v c_{\beta} \left( 1  + \zeta_{l} (1-  y_{\tau} \mu t_{\beta} /A)\right), \\
\zeta_{l}  &=& \frac{2 \alpha }{ \pi y_{\tau} } m_{\widetilde{\gamma}} A
 F(\widetilde{m}_{f}, \widetilde{m}_{f},m_{\widetilde{\gamma}}),
\end{eqnarray}
and $\gamma_{\tau}$ is a loop factor that 
has a simple expression
in the mass degenerate sfermion case,
\begin{equation}
\gamma_{\tau} =\frac{\sigma \zeta_{l} }{\left( 1  + \zeta_{l} (1- y_{\tau} \mu t_{\beta}/A )\right)}
\approx \sigma \zeta_{l}. 
\end{equation}
 As in the down-type quark sector, $\gamma_{t}$ encodes the dependence on the SUSY spectra and
parametrizes the breaking of the alignment 
between soft trilinear and Yukawa sectors caused by the tree level component
to the tau lepton mass.
Although not diagonal in the gauge basis, the matrix ${\bf m}_{L}$ 
can be brought to diagonal form in the mass basis by a biunitary diagonalization,
$ ({\cal V}^{l}_{L})^{\dagger} {\bf m}_{L} {\cal V}^{l}_{R}
=  \left( m_{e}, m_{\mu}, m_{\tau} \right)$.
The lepton mass matrix given by Eq.~\ref{mLmat}
makes the following predictions for the lepton mass ratios,
\begin{eqnarray}
\frac{m_{e}}{m_{\mu}} &=& \frac{1}{9} \lambda^{2} ( 1 -  \frac{2}{9} \lambda^{2} + \frac{5}{3} \gamma_{\tau} \lambda ) +
 {\cal O}(\lambda^{4}) 
 , 
 \\ \frac{m_{\mu}}{m_{\tau}} &=&  3 \gamma_{\tau} \lambda 
 ( 1 + \frac{1}{9} \lambda^{2} - \frac{1}{3} \gamma_{\tau} \lambda)
  + {\cal O}(\lambda^{3}).
 \label{leptonrats}
 \end{eqnarray}
We can relate $\lambda$ and $\gamma_{\tau}$
with dimensionless and approximately renormalization scale independent
fermion mass ratios; to first order,
\begin{equation}
\lambda = 3 \left( \frac{m_{e}}{m_{\mu}}\right)^{1/2}, \quad 
\gamma_{\tau} = \frac{1}{9} \left(  \frac{m_{\mu}^{3}}{m_{\tau}^{2} m_{e}} \right)^{1/2}.
\end{equation}
Using the invariant running lepton mass ratios 
determined from experiment we obtain,
\begin{eqnarray}
\lambda &=& 0.206480 \pm 0.000002, 
\label{Eq:lambdal} \\
\gamma_{\tau} &=&  0.09495 \pm 0.0001, 
\label{Eq:gammatau}
\end{eqnarray}
Interestingly, these values of $\lambda$ and $\gamma_{\tau}$ 
are consistent with the values of $\lambda$ and $\gamma_{b}$ 
determined in the down-type quark sector. This 
surprising coincidence unveils
two relations,
\begin{eqnarray}
\left( \frac{m_{d}}{m_{s}} \right) ^{1/2}
&\approx & 3 \left( \frac{m_{e}}{m_{\mu}} \right)^{1/2}, 
\\
\left( \frac{m_{s}^{3}}{m_{b}^{2}m_{d}} \right) ^{1/2}
&\approx&  \frac{1}{9} \left( \frac{m_{\mu}^{3}}{m_{\tau}^{2}m_{e}} \right)^{1/2}. 
\end{eqnarray}
These
mass relations may be considered experimental evidence 
supporting the consistency of this scenario. 

To explain the origin of the factor '{\bf 3}' in the entry (22) of the lepton mass matrix,
let us assume that matter superfields of different families 
group as usual in the representations $\overline{\bf 5}_{a}$ and ${\bf 10}_{a}$ of $SU(5)$ 
($a=1,2,3$), 
while the Higgs superfields, ${\cal H}_{u}$ and ${\cal H}_{d}$, 
belong to the representations ${\bf 5}$ and $\overline{\bf 5}$ respectively. 
If the flavor symmetric tensor superfield 
${\cal S}$ transforms as the representation ${\bf 75}$
of $SU(5)$, then the operator,
\begin{equation}
\frac{1}{M} {\cal S}^{ab} ({\bf 75}) {\cal H}_{d} {\bf 10}_{a} {\overline{\bf 5}}_{b} \neq 0
\end{equation}
will generate the entries (11) and (22) in the lepton and down-type quark 
mass matrices. Since the $SU(5)$ tensor product ${\cal S} {\cal H}_{d}$ includes the 
representation $\overline{\bf 45}$ of $SU(5)$,
it will generate the additional factor '{\bf 3}' of Georgi \& Jarlskog 
\cite{Georgi:1979df,bhrr} 
in the lepton mass matrix. Furthermore, the flavor antisymmetric tensor ${\cal A}$
must transform as a $SU(5)$ singlet; then the operator, 
\begin{equation}
\frac{1}{M} {\cal A}^{ab} ({\bf 1}) {\cal H}_{d} {\bf 10}_{a} {\overline{\bf 5}}_{b} \neq 0
\end{equation}
will generate correctly the entries (12) and (21)
in the down-type quark and lepton mass matrices. Finally, the ${\rm U(2)}_{H}$ flavor
vector superfield ${\cal F}$ could transform as a singlet or under the 
representation ${\bf 24}$ of $SU(5)$; then the operator
\begin{equation}
{\cal F} ({\bf 1,24}) \rightarrow 
\frac{1}{M} {\cal F}^{a} {\cal H}_{d} {\bf 10}_{a} {\overline{\bf 5}}_{3} \neq 0
\end{equation}
would generate the entries (a3) ($a=1,2$), and analogously the entries (3a) 
from the operator 
${\cal F}^{a} {\cal H}_{d} {\bf 10}_{3} {\overline{\bf 5}}_{a}$. 
\subsection{The charm-quark mass problem}
Assigning the ${\rm U(2)}_{H}$--flavor fields ${\cal S}$, ${\cal A}$ and
${\cal F}$ to the representations ${\bf 75}$, ${\bf 1}$ and ${\bf 1}$
of $SU(5)$ respectively 
implies that two of the associated operators in the up--type quark 
sector are exactly zero,
\begin{eqnarray}  
\frac{1}{M} {\cal S}^{ab} ({\bf 75}) {\cal H}_{u} {\bf 10}_{a} {\bf 10}_{b} = 0 \\
\frac{1}{M} {\cal A}^{ab}({\bf 1}) {\cal H}_{u} {\bf 10}_{a} {\bf 10}_{b} = 0
\end{eqnarray}
where $a,b=1,2$. If this were the case the up-type quark soft trilinear matrix would
be of the form,
\begin{equation}
{\bf A}_{U} = A
\left[
 \begin{array}{ccc}
 0 & 0  & \sigma \lambda^{2} \\
0  &  0 &  \sigma \lambda \\
\sigma \lambda^{2} & \sigma \lambda & 1
\end{array}
\right],
\label{AUmat}
\end{equation}
implying that the up--quark mass is masless and
the charm mass is, to first order, 
$m_{c} \approx \gamma_{t}^{2} \lambda^{2} m_{t}$. 
This is inconsistent with the values for $\gamma_{b}$
and $\gamma_{\tau}$ required by phenomenology
in the down-type quark and lepton sectors respectively.

We propose two possible solutions to fix the
charm quark mass problem .
One is to extend the ${\rm U(2)}_{H}$ flavor-breaking sector.
Let us assume that there are two sets of ${\rm U(2)}_{H}$ flavor-breaking fields,
one set transforming as a $SU(5)$ singlet and the other transforming 
as a ${\bf 75}$ under $SU(5)$,
\begin{eqnarray}
\left ( {\cal S}({\bf 75}) , {\cal A}({\bf 75}),
{\cal F}({\bf 75}) \right) \\
\left ( {\cal S}({\bf 1})  , {\cal A}({\bf 1}) ,
 {\cal F}({\bf 1}) \right). 
\end{eqnarray}
Let us assume that 
${\cal S}({\bf 75})$,  ${\cal A}({\bf 1})$,
and ${\cal F}({\bf 1})$ get vacuum expectacion values
as described in Eqs.~\ref{Svev}--\ref{Fvev} correctly generating
the down--type quark and lepton matrices and also the entries (3a) 
and (a3) in the up-type quark mass matrix. Additionally we assume that 
${\cal S}({\bf 1})$, gets a vev 
\begin{equation}
<{\cal S}({\bf 1})> =\left(
 \begin{array}{cc}
0  &  0 \\
 0 & {\cal V}_{\cal A}
\end{array}
\right) \theta^{2},
\end{equation}
where ${\cal V}_{\cal A}  = \lambda ^{2}  M_{\rm F} \widetilde{m}$.
This vev will generate an entry (22) in the up--type quark mass matrix, 
of the correct size to explain the charm quark mass, through the
operator 
\begin{equation}  
\frac{1}{M} {\cal S}^{ab} ({\bf 1}) {\cal H}_{u} {\bf 10}_{a} {\bf 10}_{b} \neq 0. 
\end{equation}
A second possibility is to use 
non-renormalizable operators with higher order powers
of $1/M_{\rm F}$ \cite{bhrr}. In this case we need to introduce 
at least one new ${\rm U(2)}_{H}$--flavor singlet scalar field, $\Sigma$. 
If $\Sigma$ transforms as a representation {\bf 24} under $SU(5)$,
we could generate the entries (22), (12) and (21) in the up-quark mass matrix from 
the operators,
\begin{eqnarray}
\frac{1}{M^{2}} \Sigma ({\bf 24}) {\cal S}^{ab} ({\bf 75}) {\cal H}_{u} {\bf 10}_{a} {{\bf 10}}_{b} \neq 0 \\
\frac{1}{M^{2}} \Sigma ({\bf 24}){\cal A}^{ab} ({\bf 1}) {\cal H}_{u} {\bf 10}_{a} {{\bf 10}}_{b} \neq 0 
\end{eqnarray}
If $\left<\Sigma \right> = \lambda M_{\rm F}$, we would generate an entry (22) 
of order $\gamma_{t}\lambda^{2}$ and 
entries (12) and (21) of order $\gamma_{t}\lambda^{3}$
in the up-type quark mass matrix.
Unfortunately, a (12)-(21) entry of order $\gamma_{t}\lambda^{3}$ would require
a $\gamma_{t}$ inconsistent with the value of $\gamma_{b}$ calculated in 
the down-type sector and with $\gamma_{t}$ being a loop factor. 
One way to save the higher order mechanism would
be to keep the up quark massless, 
removing the $\Sigma ({\bf 24})\, {\cal A} ({\bf 1}) \, {\cal H}_{u} \, {\bf 10}\, {\bf 10}$ 
term by imposing an additional
discrete symmetry. For instance, a $Z_{2}$ symmetry under which ${\cal A}$ and  
the lepton and down-type quark right handed fields have parity $(-)$ and the rest
of superfields have parity $(+)$ would do the job.

In these two cases, an up-type quark soft trilinear matrix would
be generated of the form,
\begin{equation}
{\bf A}_{U} =  A
\left[  \begin{array}{ccc}
 0 & 0  & \sigma \lambda^{2} \\
0  &  \sigma \lambda^{2} &  \sigma \lambda \\
 \sigma \lambda^{2} & \sigma \lambda & 1
\end{array}
\right].
\end{equation}
This matrix would account correctly for the charm/top quark mass ratio,
consistently with the 
values of $\gamma_{b}$ and $\gamma_{\tau}$
calculated in previous subsections.
Unfortunately it would predict an up-quark mass, $m_{u} \approx \gamma^{2} \lambda^{4}$,
one order of magnitude heavier than the measured value.
Therefore we are forced to prevent the flavor breaking 
field ${\cal F}(\bf 1)$ from mixing with the
up-type Yukawa operators. This could be enforced by impossing an aditional
discrete or U(1) symmetry. As a consequence we will obtain,
\begin{equation}
{\bf A}_{U} =  A
\left[  \begin{array}{ccc}
 0 & 0  & 0 \\
0  &  \sigma \lambda^{2} & 0 \\
0 & 0 & 1
\end{array}
\right].
\end{equation}
Aditionally we could assume that the flavor field ${\cal F}(\bf 75)$ gets a vev of the form,
\begin{equation}
<{\cal F}(\bf 75) >  =\left(
 \begin{array}{c}
0 \\
-{\cal V}_{\cal F}
\end{array}
\right) \theta^{2},
\label{F75vev}
\end{equation}
where ${\cal V}_{\cal F}=  \lambda M_{\rm F} \widetilde{m}$.
This would generate entries (23) and (32) in ${\bf A}_{U}$ of the form,
\begin{equation}
{\bf A}_{U} =  A
\left[  \begin{array}{ccc}
 0 & 0  & 0 \\
0  &  \sigma \lambda^{2} &  -\sigma \lambda \\
0 &  - \sigma \lambda & 1
\end{array}
\right].
\end{equation}
These two solutions are both phenomenologically viable
as we will see in more detail below.
\subsection{The up-quark mass problem}
The extension of the ${\rm U(2)}_{H}$ flavor-breaking sector allows us 
to correctly generate the charm quark mass. 
On the other hand, the up quark still remains massless.
Although the possibility of a massless up quark has been considered in the past
as a solution to the strong CP problem, 
more recent studies of pseudoescalar masses and decay constants, along with other arguments,
strongly suggest that the up quark mass is non-zero \cite{upmassless,Hagiwara:fs}.
We can easily generate the up quark mass in the scenario with two sets of
flavor-breaking superfields through a small perturbation 
of the vev of the ${\cal S}(\bf 1)$ field. Let us assume that,
\begin{equation}
<{\cal S}({\bf 1})> =\left(
 \begin{array}{cc}
u  &  0 \\
 0 & {\cal V}_{\cal A}
\end{array}
\right) \theta^{2},
\end{equation}
where $\left( u,{\cal V}_{\cal A}\right) =
\left( \lambda ^{6}, \lambda^{2} \right) M \widetilde{m}$.
Alternatively, we could generate an up--quark mass in the scenario
with higher order operators in $1/M$ if we perturb the vev
of the ${\cal S}(\bf 75)$ field in the form,
\begin{equation}
<{\cal S}({\bf 75})> =\left(
 \begin{array}{cc}
w  &  0 \\
 0 & {\cal V}_{\cal F}
\end{array}
\right) \theta^{2},
\end{equation}
where $\left( w,{\cal V}_{\cal F}\right) =
\left( \lambda ^{5}, \lambda \right) M \widetilde{m}$.
We observe that this perturbation of the ${\cal S}(\bf 1)$ or ${\cal S}(\bf 75)$
vevs does not affect the predictions in the down-type quark and lepton sectors. 

In these two cases an up-type quark soft trilinear matrix would
be generated of the form,
\begin{equation}
{\bf A}_{U} =  A
\left[  \begin{array}{ccc}
 \sigma \lambda^{6} & 0  & 0 \\
0  &  \sigma \lambda^{2} &  0 \\
 0 & 0 & 1
\end{array}
\right].
\label{AUsol1}
\end{equation}
or alternativelly if we assume that the field ${\cal F}(\bf 75)$ gets a vev
as in Eq.~\ref{F75vev},
\begin{equation}
{\bf A}_{U} =  A
\left[  \begin{array}{ccc}
 \sigma \lambda^{6} & 0  & 0 \\
0  &  \sigma \lambda^{2} &  -\sigma \lambda \\
 0 & -\sigma \lambda & 1
\end{array}
\right].
\label{AUsol2}
\end{equation}
These textures can both correctly account for the up and charm to top 
quark mass ratios as we will see in the next subsection. 
\subsection{Up-type quark masses and CKM predictions}
In the first case considered in Eq.~\ref{AUsol1}
we obtain a simple expression for the radiatively corrected up--type quark mass matrix,
\begin{equation}
{\bf m}_{U} =  \widehat{m}_{t}
  \left[
 \begin{array}{ccc}
 \gamma_{t} \lambda^{6}  & 0 &  0  \\
 0 &  \gamma_{t} \lambda^{2} &  0 \\
0 & 0 &  1
\end{array}
\right],
\label{murad1}
\end{equation}
where $\widehat{m}_{t}$ is the running top quark mass
defined by,
\begin{eqnarray}
\widehat{m}_{t} &=& y_{t} v s_{\beta} \left( 1  + \zeta_{u} (1-  y_{t} \mu / (A t_{\beta}))\right), \\
\zeta_{u}  &=& 
\frac{2 \alpha_{s} }{3 \pi y_{t} } m_{\widetilde{g}} A
 F(\widetilde{m}_{f}, \widetilde{m}_{f},m_{\widetilde{g}}),
\label{mtgt}
\end{eqnarray}
and $\gamma_{t}$ is loop factor given by,
\begin{equation}
\gamma_{t} = \frac{\zeta_{u}  \sigma }{\left( 1  + \zeta_{u} (1- y_{t} \mu /(t_{\beta}A) )\right)}, 
\approx \zeta_{u} \sigma
\end{equation}
analogous to the ones 
defined in the down-type quark and lepton sectors.
In the second case considered in Eq.~\ref{AUsol2}
the radiatively corrected up--type quark mass matrix
is given by,
\begin{equation}
{\bf m}_{U} =  \widehat{m}_{t}
  \left[
 \begin{array}{ccc}
 \gamma_{t} \lambda^{6}  & 0 &  0  \\
 0 &  \gamma_{t} \lambda^{2} &  -\gamma_{t} \lambda \\
0 & -\gamma_{t}\lambda &  1
\end{array}
\right],
\label{murad2}
\end{equation}
The phenomenological implications of a mass matrix of a form 
similar to Eq.~\ref{murad2} have been
studied by one of the authors in Ref.~\cite{Ferrandis:2004ng}. 
Although not diagonal in the gauge basis, the matrix ${\bf m}_{U}$ 
can be brought to diagonal form in the mass basis by a unitary diagonalization,
$ ({\cal V}^{u}_{L})^{\dagger} {\bf m}_{U} {\cal V}^{u}_{R}
=  \left( m_{u}, m_{c}, m_{t} \right)$.
It makes the following predictions for the up-type quark mass ratios,
\begin{eqnarray}
\frac{m_{u}}{m_{c}} &=&\lambda^{4}  ( 1 + \gamma_{t}( 1 + \gamma_{t}) ) + 
{\cal O}(\gamma^{2} \lambda^{6}), 
\label{upquarkrats1}\\ 
\frac{m_{c}}{m_{t}} &=&  \gamma_{t} \lambda^{2} (1- \gamma_{t}) ( 1 - 2 \gamma_{t}^{2}  \lambda^{2} )
  + {\cal O}(\lambda^6).
 \label{upquarkrats2}
 \end{eqnarray}
In both cases we can express $\lambda$ and $\gamma_{t}$
as a function of up--type quark mass ratios, to first order, as,
\begin{equation}
\lambda = \left( \frac{m_{u}}{m_{c}}\right)^{1/4}, \quad 
\gamma_{t} = \left(  \frac{m_{c}^{3}}{m_{t}^{2} m_{u}} \right)^{1/2}.
\end{equation}
Using the invariant running quark mass ratios 
determined from experiment (see appendix) 
and Eqs.~\ref{upquarkrats1}--\ref{upquarkrats2}
we can determine $\lambda$ and $\gamma_{t}$
numerically to be,
\begin{eqnarray}
\lambda &=& 0.225 \pm 0.015, 
\label{Eq:lambdau} \\
\gamma_{t} &=&  0.071 \pm 0.019. 
\label{Eq:gammat}
\end{eqnarray}
These values for $\lambda$ and $\gamma_{t}$ coincide
with the values for $\lambda$, $\gamma_{b}$, and 
$\gamma_{\tau}$ determined from measured fermion masses
in the down-type quark and lepton sectors. 
This surprising coincidence unveils
two more relations between dimensionless fermion
mass ratios,
\begin{eqnarray}
\left( \frac{m_{d}}{m_{s}} \right) ^{1/2} &
\approx  & \left( \frac{m_{u}}{m_{c}} \right) ^{1/4}  \approx
3 \left( \frac{m_{e}}{m_{\mu}} \right)^{1/2}, \\
\left( \frac{m_{s}^{3}}{m_{b}^{2}m_{d}} \right) ^{1/2}
&\approx&  
\left( \frac{m_{c}^{3}}{m_{t}^{2}m_{u}} \right) ^{1/2} \approx
\frac{1}{9} \left( \frac{m_{\mu}^{3}}{m_{\tau}^{2}m_{e}} \right)^{1/2}. 
\end{eqnarray}
The up-type quark diagonalization matrix can be calculated 
as a function of $\lambda$ and $\gamma_{t}$.
The up--quark diagonalization matrix can be used in combination with the
down-quark diagonalization matrix to obtain an expression
for the CKM mixing matrix, $ {\cal V}_{CKM} = ({\cal V}^{u}_{L})^{\dagger}{\cal V}^{d}_{L}$.
We are considering two cases: in the first one $A_{D}$ is given by
Eq.~\ref{Asoft} and $A_{U}$ is given by Eq.~\ref{AUsol2}, in the second case
$(A_{D})_{12}$ gets an additional factor 2 from the ${\cal F}(\bf 1)$ 
vev in Eq.~\ref{vavf2}
while $A_{U}$ is given by Eq.~\ref{AUsol1}. In both cases
using that $\gamma = \gamma_{b} \approx \gamma_{t}$ 
we obtain the same prediction for $\left| {\cal V}_{CKM}\right|^{\rm{theo}}$
to leading order in $\lambda$, 
\begin{equation}
\left[
 \begin{array}{ccc}
1 - \frac{1}{2} \lambda^{2}  &  \lambda   & 
 \gamma \lambda^{2}  \\
\lambda &  1 - \frac{1}{2} \lambda^{2} \left(1 + 2 \gamma^{2}\right) & 
 2 \gamma \lambda
 \\
 \gamma \lambda^{2} 
 & 2 \gamma \lambda
 & 1 - 2 \gamma^{2} \lambda^{2}
\end{array}
\right].
\end{equation}
Using the experimentally determined values for $\lambda$, $\gamma_{b}$, and
$\gamma_{t}$ in Eqs.~\ref{Eq:lambdad},  \ref{Eq:gammab} \& \ref{Eq:gammat}
we obtain the following numerical 
theoretical prediction for the $\left| {\cal V}_{CKM} \right|_{\rm{theo}}$ 
elements,
\begin{equation}
\left[
 \begin{array}{ccc}
0.976\pm 0.008 & 0.216\pm 0.035 & 0.0039 \pm 0.0006 \\ 
0.216\pm 0.035 & 0.974 \pm 0.007 & 0.035 \pm 0.006 \\
0.0039 \pm 0.0006 & 0.035 \pm 0.006  & 0.9993 \pm 0.0001
\end{array}
\right]. 
\end{equation}
If we now compare  $\left| {\cal V}_{CKM} \right|_{\rm{theo}}$ 
with the 90~\% C.L. 
experimental compilation of CKM matrix elements from the 
PDG compilation (see appendix),  
we observe that  $\left| {\cal V}_{CKM} \right|_{\rm{theo}}$ 
accounts perfectly for the measured flavor violation in the Standard Model. 
There is now very good agreement with 
the experimental data on the entry $\left|V_{cb}\right|$.
We also obtain the same prediction for $\left|V_{td}\right|$,
$\left|V_{td}\right| \approx \left|V_{ub}\right|$ 
The flavor violation in the upper left sector of the up-type quark
mass matrix is not constrained by the CKM matrix,
since the flavor violation in the up--type quark sector 
does not affect the entries $(12)$ and $(13)$ to leading order in $\lambda$. 
\subsection{Higher order Yukawa couplings}
Yukawa couplings which are not generated at one loop could be generated
at higher orders. For instance, the Yukawa coupling $({\bf Y}_{U})_{13}$
could be generated at two loops through a diagram with gluino and
Higgs exchange and three soft trilinear vertices:
$({\bf A}_{D})_{12}$, $({\bf A}_{D})_{22}$ and $({\bf A}_{U})_{23}$ \cite{radovan}. 
We are interested in an overestimation of this 2-loop Yukawa
coupling. Assuming that all the sparticles in the loop have masses
of the same order, to maximize the loop factor, we obtain
\begin{equation}
({\bf Y}_{U})_{13}^{\rm 2-loop} \simeq \left( \frac{2 \alpha_{s}}{3 \pi} \right)
\left( \frac{1}{4 \pi} \right)^{2} \left( \frac{v}{m_{\widetilde{q}}}\right)^{2}
c^{2}_{\beta} \lambda^{4},
\end{equation}
here $v=175$~GeV. The ratio $v/m_{\widetilde{q}}$, the $c_{\beta}$ factors 
($c_{\beta} =\cos \beta$)  
and the $\lambda$ factors come from the three
soft trilinear vertices. To facilitate the comparison with the one-loop
generated Yukawa couplings we will express this in powers of $\lambda$.
Using that $\lambda \simeq 0.2$
and $\gamma \simeq 0.1$, we obtain,
\begin{equation}
({\bf Y}_{U})_{13}^{\rm 2-loop} \simeq 
\frac{\gamma \lambda^{10}}{\tan^{2}\beta} 
\left( \frac{1~{\rm TeV}}{m_{\widetilde{q}}}\right)^{2}.
\label{YUsup}
\end{equation}
We note that this 2-loop generated Yukawa couplings is 
very suppressed when compared with the one-loop generated couplings
and for all practical purposes it can be considered zero.
\section{Suppression of flavor changing processes by radiative alignment}
It has been pointed out recently \cite{Ferrandis:2004ng} 
that the radiative generation of fermion
masses through flavor violation in the soft breaking terms may 
allows us to overcome the present experimental constraints on 
some of the supersymmetric contributions to flavor changing processes
more easily than other flavor models.
This is a necessary requirement for the consistency 
of any supersymmetric model \cite{susyFC}.
Correlations between radiative mass generation and dipole operator phenomenology
were first pointed out in Ref.~\cite{Kagan:1994qg}. In this scenario,
as a consequence of the approximate radiative alignment 
between Yukawa and soft trilinear matrices there is
an extra suppression of the supersymmetric contributions to flavor changing 
processes coming from the soft trilinear sector. 
For calculational purposes
it is convenient to rotate the squarks to the so-called superKM basis, the
basis where gaugino vertices are flavor diagonal ~\cite{masieroFCNC}. 
In this basis, the entries in the soft trilinear matrices are directly proportional
to their respective 
contributions to flavor violating proceses. For instance,
the soft trilinear matrix ${\bf A_{D}}$
in the superKM basis as given by,
\begin{equation}
 {\bf A}_{D}^{\rm{SKM}} =
({\cal V}^{d}_{L})^{\dagger} {\bf A}_{D} {\cal V}^{d}_{R}.
\end{equation}
Assuming the soft trilinear matrix ${\bf A_{D}}$ is given by Eq.~\ref{Asoft},
which corresponds to the proposed solution where
one half of $\left| V_{cb}\right|$ is generated in the up-sector,
and the diagonalization matrices ${\cal V}^{d}_{L,R}$ are calculated
from Eq.~\ref{YukDrad}, we obtain, to leading order in 
$\lambda$ and $\gamma_{b}$,
\begin{equation}
 {\bf A}_{D}^{\rm{SKM}} =
 A_{b}
\left[
 \begin{array}{ccc}
-\sigma \lambda^{3} &  \sigma \theta \lambda^{5} & \sigma \lambda^{4} \\
2\sigma \gamma_{b} \lambda^{3} &  \sigma\lambda &   (\gamma_{b}-\sigma) \lambda  \\
2\sigma \lambda^{2} & (\gamma_{b}-\sigma) \lambda  &  1 + 2\sigma \gamma_{b}\lambda^{2}
\end{array}
\right].
\end{equation}
Here we will assume that $\sigma$, as defined in Eq.~\ref{sigma}, is 1.
We observe that the entry $(21)$ is suppressed by an additional factor
$\gamma_{b} \lambda$
as a consequence of the radiative alignment between Yukawa and soft trilinear matrices.
Moreover in the SKM basis the left handed down-type squark 
soft mass matrix is given by, 
\begin{equation}
 (\widetilde{{\bf M}}^{2}_{D})_{LL}^{\rm{SKM}} =
({\cal V}^{d}_{L})^{\dagger}  (\widetilde{{\bf M}}^{2}_{D})_{LL} {\cal V}^{d}_{L},
\end{equation}
and analogously for the right handed soft mass matrix.
Assuming the soft trilinear texture from
Eq.~\ref{Msoft} we obtain for $({\bf M}^{2}_{D})_{LL}^{\rm{SKM}}$,
to leading order in $\lambda$ and $\gamma_{b}$,
\begin{equation}
m_{\widetilde{b}_{L}}^{2}  \left[
 \begin{array}{ccc}
1 + 2 \sigma^{\prime 2}\lambda^{4} &   -  \sigma^{\prime 2}\lambda^{3} &  \sigma^{\prime 2}\lambda^{4} \\
- \sigma^{\prime 2} \lambda^{3} &  1+ 2   \sigma^{\prime 2}\lambda^{2} &  -  \sigma^{\prime 2}\lambda  \\
-  \sigma^{\prime 2}\lambda^{4} & - \sigma^{\prime 2}\lambda  &  (1 + 2  \sigma^{\prime 2}\gamma_{b} \lambda^{2})
\end{array}
\right].
\end{equation}
Here $\sigma^{\prime}$ was defined in Eq.~\ref{sigmap}.
We note in the limit $\sigma^{\prime} \ll 1$ the contributions
from soft mass matrices to flavor changing processes would be suppressed. 
The following results would not change much assuming instead the second
solution we proposed in the previous section. 
Using the values given by Eqs.~\ref{Eq:lambdad} \& \ref{Eq:gammab}
for $\lambda$ and $\gamma_{b}$ 
the amount of soft flavor violation required to 
fit quark masses and mixing angles is determined.

The entry most constrained experimentally
in the soft mass matrices is the entry (12), which in our
scenario gives the dominant contribution 
to the $K_{L}$--$K_{S}$ mass difference.
If we assume that $\sigma^{\prime}=1$, {\it} i.e. $A_{b}=m_{\widetilde{q}}$,
then we obtain that 
the squark spectra must be $m_{\widetilde{q}}> 2~$TeV
to avoid the saturation of the experimental measurement,
$\Delta m_{K} = (3.490 \pm 0.006)\times 10^{-12}$~MeV \cite{Hagiwara:fs}.
This constraint is considerably 
milder if the gluino-squark mass ratio
is much larger or smaller than one.
On the other hand if $A_{b}= m_{\widetilde{q}}/2$ the flavor violating
soft mass matrices receive an additional $1/4$ factor through the
$\sigma^{\prime}$ coefficient defined in Eq.~\ref{sigmap}.
In this case we would saturate the experimental measurement for
$m_{\widetilde{q}}> 400~$GeV and we would predict 
$\Delta m_{K} < 7 \times 10^{-13}$ for $m_{\widetilde{q}}> 1~$TeV.
Furthermore, in the limit $\sigma^{\prime} \ll 1$ this contribution 
goes to zero and the soft trilinear contribution dominates.
The soft trilinear contribution becuase of the extra suppression factor
$\gamma_{b} \lambda$ is very suppressed. Assuming a large value
of $\tan\beta$, $m_{\widetilde{q}}>700$~GeV and any gluino-squark
mass ratio it generates a contribution
to $\Delta m_{K}$ below the experimental uncertainity.

The entry (13) in the soft mass matrix 
$(\widetilde{{\bf M}}^{2}_{D})_{LL}^{\rm{SKM}} $ gives also 
the dominant contribution to  $\Delta m_{B}$. 
We note that the entry (13) appears as a consequence
of the possible mixing between flavor breaking and
flavor singlet susy breaking fields through operators of the form,
\begin{equation}
\frac{1}{M M_{\rm F}}
\int  d^{4} \theta  \rho \left( ( \kappa^{\prime }{\cal G}^{\dagger}+ \eta^{\prime} {\cal J}^{\dagger} )
 {\cal F}^{b} \phi^{\dagger} \Psi_{b}  + {\rm h.c.} \right)
\label{softmix}
\end{equation}
The texture under consideration predicts a contribution to $\Delta m_{B}$ 
for $m_{\widetilde{q}} > 600~$GeV and any gluino mass ratio
which is below uncertainity of the experimental measurement
$\Delta m_{B} = (3.22 \pm 0.05)\times 10^{-10}$~MeV.
We note that this constraint can be 
avoided if operators of the form in Eq.~\ref{softmix}
are not allowed by the underlying supersymmetric theory,
{\it i.e.} if we assume that $\rho=0$.

Finally, from the measured $b \rightarrow s \gamma$ decay rate,
one can obtain limits on the entry (23) \cite{Bertolini:1986tg}.
In general the flavor violating soft trilinear gives the dominant contribution
to this process. 
Assuming a large value of $\tan\beta$, $\tan\beta >30$, $A_{b}\simeq m_{\widetilde{q}}$,
and $m_{\widetilde{q}}> 500~$GeV we obtain 
a contribution to $ B(b\rightarrow s \gamma)  < 3.4 \times 10^{-5}$,
which is still below the uncertainty of the experimental measurement,
$B(b\rightarrow s \gamma) = (3.3 \pm 0.4)\times 10^{-4}$.
Using known expressions \cite{masieroFCNC,Bertolini:1986tg}
we can also calculate the soft mass, {\it i.e.} LL, contribution 
to $B( b\rightarrow s \gamma)$.
This is in general suppressed when compared with the LR, {\it i.e.}
soft trilinear contribution, by a factor,
\begin{equation}
\frac{1}{6} \left( \frac{m_{b}}{m_{\widetilde{g}}} \right) \frac{(\delta_{12}^{d})_{LL}}{(\delta_{12}^{d})_{LR}} 
\approx  3 \times 10^{-3} t_{\beta} \left(\frac{m_{\widetilde{b}}}{m_{\widetilde{g}}}\right)
\end{equation}
where we used that, 
\begin{eqnarray}
(\delta_{12}^{d})_{LR} &=&  (\gamma_{b}-1)\lambda \left( \frac{A_{b}}{m_{\widetilde{b}}}
\right)\left( \frac{v}{m_{\widetilde{b}}}\right) \frac{1}{t_{\beta}} \\
(\delta_{12}^{d})_{LL} &=&  \lambda
\end{eqnarray}
and $v$ and $\gamma_{\tau}$ are given by 
$v=175$~GeV and $\gamma_{b}\approx0.1$. 
 Even considering very large $\tan\beta$ values the LL contribution in this model
 is one order of magnitude smaller than the LR contribution
 to this process. Therefore the approximate constraints on the supersymmetric spectra 
calculated above from the LR contribution to $B( b\rightarrow s \gamma)$, 
while ignoring the LL contributions, are still valid.

We can perform a similar analysis of flavor
changing processes in the lepton sector.
As in the squark sector
it is convenient for calculational purposes,
to rotate the sleptons to a basis
where gaugino vertices are flavor diagonal. 
The soft trilinear matrix ${\bf A_{L}}$ in the superKM basis is given by,
\begin{equation}
 {\bf A}_{L}^{\rm{SKM}} =
({\cal V}^{l}_{L})^{\dagger} {\bf A}_{L} {\cal V}^{l}_{R}
\end{equation}
Assuming the soft trilinear texture from
Eq.~\ref{ALsoft} with $\sigma=1$ we obtain,
to leading order in $\lambda$ and $\gamma_{\tau}$,
\begin{equation}
 {\bf A}_{L}^{\rm{SKM}} =
A_{\tau}  \left[
 \begin{array}{ccc}
\frac{1}{3} \lambda^{3} &   \frac{2}{3} \gamma_{\tau} \lambda^{3} & \frac{2}{3}  \lambda^{2} \\
\frac{4}{3} \gamma_{\tau}\lambda^{3} &   3 \lambda &   (\gamma_{l}-1) \lambda  \\
\frac{4}{3} \lambda^{2} &  (\gamma_{l}-1) \lambda  &  (1 + 2 \gamma_{\tau}\lambda^{2})
\end{array}
\right].
\end{equation}
Using the values given by Eqs.~\ref{Eq:lambdal} \& \ref{Eq:gammatau}
for $\lambda$ and $\gamma_{\tau}$, 
the amount of soft lepton flavor violation is determined.
The entry (12) contributes to $B(\mu \rightarrow e \gamma)$,
which is the most experimentally constrained lepton flavor violating process.
For the texture under consideration, 
assuming a large value of $\tan\beta$, $\tan\beta \approx 50$, $A_{b} \simeq m_{\widetilde{l}}$,
a photino lighter than the sleptons and
$m_{\widetilde{l}}> 1~$TeV we obtain 
a branching fraction 
$\Gamma_{\mu \rightarrow e \gamma} < 8 \times 10^{-12}$,
which is still below the current experimental limit,
$\Gamma^{\rm{exp}}_{\mu \rightarrow e \gamma} 
<  1.2 \times 10^{-11}$.
The predictions for $\Gamma_{\tau \rightarrow e \gamma}$ and
$\Gamma_{\tau \rightarrow \mu \gamma}$ are proportional to the
entries (13) and (23) respectively in the soft trilinear matrix. 
For the texture under consideration, 
using the same parameter space limits, we obtain,
$\Gamma_{\tau \rightarrow e \gamma} <  10^{-10}$
and $\Gamma_{\tau \rightarrow \mu \gamma} < 2 \times 10^{-9}$; 
these two predictions are far below the experimental limits,
$\Gamma^{\rm{exp}}_{\tau \rightarrow e \gamma} 
<  2.7 \times 10^{-6}$ and $\Gamma^{\rm{exp}}_{\tau \rightarrow \mu \gamma} 
<  1.1 \times 10^{-6}$.

There are also contributions to $B(\mu\rightarrow e \gamma)$
coming from flavor violating soft masses.  
In the SKM basis the left handed charged slepton 
soft mass matrix is given by, 
\begin{equation}
 (\widetilde{{\bf M}}^{2}_{L})_{LL}^{\rm{SKM}} =
({\cal V}^{l}_{L})^{\dagger}  (\widetilde{{\bf M}}^{2}_{L})_{LL} {\cal V}^{l}_{L}
\end{equation}
Assuming the soft trilinear texture from
Eq.~\ref{ALsoft} and $\sigma^{\prime}=1$ we obtain,
to leading order in $\lambda$ and $\gamma_{\tau}$,
\begin{equation}
({\bf M}^{2}_{L})_{LL}^{\rm{SKM}} =
m_{\widetilde{\tau}_{L}}^{2}  \left[
 \begin{array}{ccc}
1 &   \frac{1}{3} \lambda^{3} & \frac{2}{3}  \lambda^{2} \\
-\frac{1}{3} \lambda^{3} &  1+ 2  \lambda^{2} &  - \lambda  \\
\frac{2}{3} \lambda^{2} & -\lambda  &  (1 + 2 \gamma_{\tau} \lambda^{2})
\end{array}
\right].
\end{equation}
The contribution of the flavor violating soft masses to $B(\mu 
\rightarrow e \gamma)$ is suppressed
compared with the contribution from the soft trilinear terms by a factor,
\begin{equation}
\frac{1}{6} 
\left( \frac{m_{\mu}}{m_{\widetilde{\gamma}}} \right) \frac{(\delta_{12}^{l})_{LL}}{(\delta_{12}^{l})_{LR}} 
\approx 5 \times 10^{-4} t_{\beta} \left(\frac{m_{\widetilde{l}}}{m_{\widetilde{\gamma}}}\right)
\end{equation}
where we used that, 
\begin{eqnarray}
(\delta_{12}^{l})_{LR} &=& \frac{2}{3} \gamma_{\tau} \lambda^{3}\left( \frac{A_{\tau}}{m_{\widetilde{l}}}
\right)\left( \frac{v}{m_{\widetilde{l}}}\right) \frac{1}{t_{\beta}} \\
(\delta_{12}^{l})_{LL} &=& \frac{1}{3} \lambda^{3}
\end{eqnarray}
and $v$ and $\gamma_{\tau}$ are given by 
$v=175$~GeV and $\gamma_{\tau} =0.95$. We note that even for a large value of $\tan\beta$
the LL contribution is much smaller than the LR contribution in this model.

To sumarize, the flavor violation present in the soft supersymmetry-breaking
sector, which is necessary in this scenario to generate fermion masses and
quark mixings radiatively, is not excluded 
by the present experimental constraints. These constraints are  
not especially stronger than in other supersymmetric flavor models. 
The approximate radiative alignment between 
radiatively generated Yukawa matrices and soft trilinear terms helps to
suppress some of the supersymmetric contributions to these processes,
especially the contribution to $B(\mu\rightarrow e \gamma)$. 
\section{Proton decay suppression}
It is generally believed that strong experimental limits on proton decay place
stringent constraints on supersymmetric grand unified models.
Nevertheless as we will see this assertion
is very dependent on the mechanism that generates
the Yukawa couplings. 
For instance, in the case of a generic 
minimal supersymmetric SU(5) model \cite{susysu5}, the superpotential,
omitting SU(5) and flavor indices, is given by 
\begin{equation}
{\cal W}_{\rm SU(5)} = 
\frac{1}{4} {\bf Y}_U\, {\bf 10}\, {\bf 10}\, {\cal H}_{u} + 
\sqrt{2} {\bf Y}_D\, {\bf 10}\, {\bf \overline{5}}\, {\cal H}_{d} +
\cdot \cdot \cdot,
\end{equation}
where $ {\bf 10}$ and ${\overline{\bf 5}}$ are matter chiral superfields
belonging to representations {\bf 10} and $\overline{{\bf 5}}$
of SU(5), respectively.
As in the supersymmetric generalization of the SM,
to generate fermion masses
we need two sets of Higgs superfields,  $ {\cal H}_{u}$
and $ {\cal H}_{d}$, belonging to representations {\bf 5} and 
$\overline{{\bf 5}}$ of SU(5).
After integrating out the colored Higgs triplet,
the presence of Yukawa couplings in the superpotential leads to effective 
dimension-five interactions which, omitting flavor indices, are of the form,
\begin{eqnarray}
{\cal W}_{\rm dim \,5} &\propto& 
\frac{1}{\rm M_{{\cal H}_{c}}} \left[\frac{1}{2}
{\bf Y}_U  {\bf Y}_{D} \left( Q Q\right) \left( Q L \right) + \right. \nonumber \\
&&\left.
{\bf Y}_U  {\bf Y}_{D} \left( U E \right) \left( U D \right) \right],
\end{eqnarray}
where ${\rm M}_{{\cal H}_{c}}$ is the coloured Higgs mass, 
and operators $\left( Q Q\right) \left( Q L \right)$ and
$\left( U E \right) \left( U D \right) $ are totally antisymmetric in color indices.
Therefore, flavor conservation in the superpotential would imply their cancellation
in the exactly supersymmetric theory,
\begin{eqnarray}
\left( Q Q \right) \left( Q L\right) &\equiv& 0, \\
\left( U E \right) \left( U D \right) &\equiv& 0.
\end{eqnarray}
In our scenario, we started by assuming that
there is a ${\rm U(2)}_{H}$ horizontal symmetry 
that guarantees the flavor conservation in 
superpotential of the supersymmetric
unified theory. 
Flavor violating couplings are only generated 
at low energy after supersymmetry-breaking. The operators
that generate flavor violation are of the form,
\begin{equation}
\frac{1}{M} ({\cal S,A}) {\cal H}_{d} 10~ \overline{5}, \quad
\frac{1}{M} {\cal S} {\cal H}_{u} 10 ~10,
\end{equation}
Integrating out the coloured higgsses we could generate in principle
baryon number violating operators of the form,
\begin{equation}
\propto {\cal S} {\cal A}  ( 10 ~\overline{5})(10 ~10),
\end{equation}
but one of our basic assumptions is that at the U(2) minimum the U(2) flavor 
breaking fields, ${\cal S}$ and ${\cal A}$, are F-terms, therefore
these operators exactly cancel and one cannot generate directly dimension five operators.
Dimension five operators could be generated at higher orders.
Since tree level interactions with coloured Higgsinos are only possible
for the third family, the generation of a dimension five proton decay
operator would require two flavor mixing couplings between first
and third generation. On the other hand,
the Yukawa coupling of the form $({\bf Y}_{U})_{13}$
is first generated at two loops and very suppressed,
as pointed out in Eqs.~\ref{YUsup}. 
As a consequence radiatively generated 
dimension five operators leading to proton decay 
are very suppressed in this scenario,
when compared with ordinary SUSY GUT predictions, which
generate flavor in the superpotential. 
Regarding the next dominant decay mode arising from dimension-six operators
via GUT gauge bosons, it has been shown that using the SuperKamiokande
limit, $\tau(p \rightarrow \pi^{0} e^{+})  > 5.3 \times 10^{33}$~years,
a lower bound on the heavy gauge boson mass, $M_{V}$, 
can be extracted, $M_{V} > 6.8 \times 10^{15}$~GeV. 
Furthermore, the proton decay rate for $M_{V} = M_{GUT}$ is far below
the detection limit that can be reached within the next years
\cite{Emmanuel-Costa:2003pu}. 
\section*{Summary}
Many recipes have been attempted to cook the observed
fermion mass hierarchies. We have shown in this
paper that a tastier dish may require the right
mix of horizontal symmetries, grand unified symmetries
and radiative mass generation. 
We have proposed an effective flavor-breaking model 
based on a U(2) horizontal symmetry which is implemented
by supersymmetry-breaking fields. As a consequence,
flavor breaking originates in the soft supersymmetry-breaking terms
and is transmitted to the Yukawa sector at low energy. 
The approximate radiative alignment between
soft trilinear matrices and the radiatively generated Yukawa matrices
at low energy helps to suppress
the supersymmetric contributions to
flavor changing processes.
The model allow us to succesfully fit the six fermion mass ratios
and the quark mixing angles with just two parameters. 
It also predicts new quantitative relations between dimensionless fermion
mass ratios in the three fermion sectors, and the quark mixing angles,
\begin{eqnarray}
\left|V_{us}\right| & \approx &
\left[ \frac{m_{d}}{m_{s}} \right] ^{\frac{1}{2}}
\approx \left[ \frac{m_{u}}{m_{c}} \right]^{\frac{1}{4}} 
\approx 3 \left[ \frac{m_{e}}{m_{\mu}} \right]^{\frac{1}{2}}, 
\\
\frac{1}{2} \left|\frac{V_{cb}}{V_{us}}\right| &\approx& \left[ \frac{m_{s}^{3}}{m_{b}^{2}m_{d}} \right] ^{\frac{1}{2}}
\approx \left[ \frac{m_{c}^{3}}{m_{t}^{2}m_{u}} \right] ^{\frac{1}{2}}
\approx  \frac{1}{9} \left[ \frac{m_{\mu}^{3}}{m_{\tau}^{2}m_{e}} \right]^{\frac{1}{2}},
\end{eqnarray}
which are confirmed by the experimental measurements. 
Moreover, the requirement of flavor conservation in the superpotential
of the grand unified theory implies the suppression of the problematic
dimension-five operators which otherwise would accelerate proton decay.
\section*{Appendix}
For the calculation of the dimensionless fermion mass ratios used in the main text, 
running fermion masses were used.
These were calculated through scaling factors including known
QCD and QED renormalization effects, which 
can be determined using known solutions to the SM RGEs.
For the charged leptons our starting point is the well known physical
masses.
For the top quark the starting point is the pole mass
from the PDG collaboration \cite{Hagiwara:fs}, 
\begin{eqnarray}
m_{t} &=& 174.3 \pm 5.1 ~\rm{GeV}.
\end{eqnarray}
For the  bottom and charm quarks the running masses, 
$m_{b}(m_{b})_{\overline{MS}}$ 
and $m_{c}(m_{c})_{\overline{MS}}$, from Refs.~\cite{bottommass} \& 
\cite{Becirevic:2001yh} are used, 
\begin{eqnarray}
m_{b}(m_{b})_{\overline{MS}} &=& 4.25 \pm 0.25 ~\rm{GeV},  \\
m_{c}(m_{c})_{\overline{MS}} &=& 1.26 \pm 0.05 ~\rm{GeV};
\end{eqnarray}
for the light quarks, u,d and s, the starting point is
the normalized $\overline{MS}$ values at $\mu=2$~GeV.
Original extractions \cite{Jamin:2001zr,Gamiz:2002nu} quoted in the literature 
have been rescaled as in \cite{Hagiwara:fs},
\begin{eqnarray}
m_{s}(2~\rm{GeV})_{\overline{MS}} &=& 117 \pm 17 ~\rm{MeV},  \\
m_{d}(2~\rm{GeV})_{\overline{MS}} &=& 5.2 \pm 0.9~\rm{MeV},\\
m_{u}(2~\rm{GeV})_{\overline{MS}} &=& 2.9 \pm 0.6 ~\rm{MeV}. 
\end{eqnarray}
For completeness we include here some functions
used in the main text. The $B_{0}$ and $F(x,y,z)$ functions, which
are used in the calculation of the one--loop finite corrections,
are given by,
\begin{equation}
B_{0} (m_{1},m_{2}) = 1 +  \ln \left(\frac{{Q}^{2}}{m_{2}^{2}}\right)
+\frac{m_{1}^{2}}{m^{2}_{2}-m_{1}^{2}} 
 \ln \left(\frac{{m}^{2}_{2}}{m_{1}^{2}}\right)
\end{equation}
where Q is the renormalization scale,
\begin{equation}
F(x,y,z) = \frac{\left[ (x^{2}y^{2} \ln \frac{y^{2}}{x^{2}} + y^{2}z^{2} \ln \frac{z^{2}}{y^{2}}
+ z^{2}x^{2} \ln \frac{x^{2}}{z^{2}} \right]}{(x^{2}-y^{2})(y^{2}-z^{2})(z^{2}-x^{2})} >0 .
\end{equation}
For completeness we also include the 90~\% C.L. 
experimental compilation of CKM matrix elements from the 
PDG compilation \cite{Hagiwara:fs},  
\begin{equation}
\left| {\cal V}_{CKM}^{\rm{exp}} \right| =   
\left[
 \begin{array}{ccc}
V_{ud} & V_{us} & V_{ub} \\
V_{cd} & V_{cs} & V_{cb} \\
V_{td} & V_{ts} & V_{tb} 
\end{array}
\right]= 
\nonumber
\end{equation}
\begin{equation}
\left[
 \begin{array}{ccc}
0.97485 \pm 0.00075 & 0.2225 \pm 0.0035 & 0.00365 \pm 0.00115 \\ 
 0.2225 \pm 0.0035  & 0.9740 \pm 0.0008 & 0.041 \pm 0.003 \\
0.0009 \pm 0.005 & 0.0405 \pm 0.0035 & 0.99915 \pm 0.00015
\end{array}
\right], 
\end{equation}
\acknowledgements
N.~Haba would like to thank the University of Hawaii for its hospitality. 
We thank Prof.~Xerxes Tata his comments. We thank
R.~Dermisek for his suggestions regarding the 
higher order generation of Yukawa couplings and proton decay operators. 
We also thank H.~Guler for many suggestions.
This work was supported by the D.O.E. grant number DE-FG03-94ER40833
and by the grants number 4039207, 14046208 \& 14740164
of the Ministry of Education and Science of Japan.


\end{document}